\newcommand{\beq}{\begin{equation}}
\newcommand{\eeq}{\end{equation}}
\documentstyle[11pt,aaspp4]{article}
\lefthead{QUILLEN AND GARNETT}
\righthead{DISK HEATING IN THE SOLAR NEIGHBORHOOD}
\begin{document}
\title{The saturation of disk heating in the solar neighborhood and 
evidence for a merger 9 Gyrs ago} 
\author{Alice C. Quillen and Donald R. Garnett} 
\affil{Steward Observatory, University of Arizona, 933 N. Cherry Ave., 
Tucson, AZ  85721 \\ E-mail: aquillen@as.arizona.edu;dgarnett@as.arizona.edu}

\begin{abstract}
We re-examine the age-velocity dispersion relation in the solar neighborhood
using improved stellar age estimates that are based on Hipparcos parallaxes
and recent stellar evolution calculations. The resulting relation shows that 
the Milky Way stellar disk was relatively quiescent, suffering little
heating or 
dispersion increase between 3 and 9 Gyrs.  However, at an age of $9\pm 1$ Gyrs 
there is an abrupt increase in the stellar velocity dispersions. To explain 
the abrupt increase (by almost a factor of 2) in dispersions we propose that 
the Milky Way suffered a minor merger 9 Gyrs ago that created the thick
disk. The quiescent phase is consistent with inefficient heating caused by 
scattering from tightly wound transient spiral structure. It 
is possible that many galaxy disks saturated radial velocity dispersions
at the point where heating from spiral structure becomes inefficient.
Surface brightness profiles
can be used to predict spiral arm winding numbers at this saturation point. 
A comparison between the observed and predicted winding 
numbers could test this hypothesis and also search for evidence of 
merger induced disk heating in nearby galaxies.

\end{abstract}

\section{The age-velocity dispersion relation in the solar neighborhood}

The observed correlation between the ages and velocity dispersions of solar
neighborhood stars has been a subject of study since the work of Spitzer 
and Schwarzschild (1951) and (1953). 
Spitzer and Schwarzschild established that scattering of
stars from initially circular orbits into more eccentric
and inclined orbits was a likely explanation for the increase in 
velocity dispersion, $\sigma$, with age $t$.
They suggested that massive gas clouds (then undetected) 
were responsible.
Molecular clouds were thought to be the sole scattering agents
(e.g., \cite{mihalas}) until \cite{lacey84_}
showed that the observed ratio of the dispersion
in the direction perpendicular to the Galactic plane
and that towards the Galactic center, ${\sigma_z / \sigma_R}$, 
was too low to be consistent with the predictions from this scattering 
process.  There was also a discrepancy between the predicted and
then measured relation between $\sigma$ and $t$: if $\sigma \propto t^\alpha$,
models predicted $\alpha \sim 0.20$, while observations of the time
yielded $\alpha \sim 0.5 $ (\cite{wielen}).
This resulted in the development of models that incorporated the
heating of the stellar disk from transient spiral structure
(\cite{barbanis}; \cite{sellwood}; \cite{carlberg}) in
addition to scattering from molecular clouds (\cite{jenkins};
\cite{jenkins92}).  

The measured value of $\alpha$ depended on stellar age estimates that 
were necessarily highly uncertain.
However, parallaxes from the Hipparcos satellite, along with improvements
in stellar evolution models for low-mass stars, have made it
possible in the past few years 
to estimate the ages of stars in the solar neighborhood
more precisely than previously possible (e.g., \cite{ng}).
Imprecise ages 
not only add scatter to a plot of stellar dispersion versus age 
but also smooth over any underlying structure in this relation 
that might actually exist in the solar neighborhood.
%
%
The theoretical work by \cite{lacey84_} and \cite{jenkins_}
was motivated in part by the smooth form of the age dispersion relation
advocated by \cite{wielen_}.   
However \cite{stromgren_} and \cite{freeman} suggested that the observations
were more consistent with little or no heating between 3 and 10 Gyrs and
an abrupt increase in the dispersion occurring at about 10 Gyrs.
\cite{freeman} suggested that disk heating mechanisms saturated
or became inefficient at $\sigma  \sim 30 $ km s$^{-1}$ and that
another process, such as a merger or large scale bar,
would be required to explain the high dispersions seen in the 
`thick disk' or in stars with ages greater than 10 Gyrs.
The thick disk was proposed to explain an excess in the number
of stars compared to predictions from an exponential disk 
model (\cite{thickdisk}; \cite{kuijken}).
However, the discreteness of this component compared to 
Population I stars in age, metallicity or dispersion
has been difficult to establish (e.g., \cite{milkywaybook}).
Stellar evolution models for the solar neighborhood 
have supported both forms for the age dispersion relation.
\cite{larsen}
found that a jump at 8 Gyrs in the age dispersion relation
was required to fit the stellar dispersions and other observational
constraints.  However, \cite{binney_}
fit the color distribution of stars observed with Hipparcos
from an evolution model with $\sigma \propto t^{0.33}$.
In this paper we use improved age estimates to re-examine 
the age-dispersion correlation.  We then compare
the resulting relations with the existing theoretical framework.

\subsection{Age and velocity component measurements for stars in the solar
neighborhood} 

We have compiled new space velocity components for solar neighborhood stars 
from the metallicity study of \cite{edvardsson93}. \cite{edvardsson93}
obtained metallicity and kinematic information for 189 nearby F and G 
star dwarf stars within 80 parsecs of the Sun. 
They  selected stars 
for metallicity and effective temperature based on Str\"omgren indices. 
Because the stars were selected 
from the magnitude-limited photometric catalogue of \cite{olsen},  
the sample is expected to be unbiased kinematically (\cite{freeman_}). 
For the stars in this sample we computed velocity components U, V, and W from
parallaxes and proper motions measured by 
the Hipparcos satellite 
and radial velocities obtained from SIMBAD.
In most cases our velocity components were within 1 km s$^{-1}$ of those
listed by \cite{edvardsson93} which were derived from ground-based
proper motions and distances.
Eight stars, (HD 17548,  18768,  68284, 78747,  98553,  144172,  155358,  
and 218504) listed by \cite{edvardsson93} have 
no published radial velocities, so for these stars we use the velocity 
components listed by \cite{edvardsson93}.
For the purposes of this
study, we excluded the eleven stars 
labeled as new spectroscopic binaries, which could introduce spurious scatter 
into the velocity dispersions. 

New ages for the \cite{edvardsson93} stars were taken from Tables 5 and 6 of \cite{ng_}. 
These ages are derived in the same manner as those of \cite{edvardsson93}, 
that is, by comparing each star's position in the HR diagram with theoretical 
stellar evolution tracks, but are based on Hipparcos parallax measurements 
and the isochrones of \cite{bert94}, as opposed to the isochrones 
from \cite{vandenberg85} used by \cite{edvardsson93}. 
The precision in derived ages depends on how far a
given star lies from the main sequence. Stars close to the main sequence
have larger uncertainties because of the smaller spacing between 
evolution tracks in the HR diagram. The uncertainty 
is generally between 0.05 and 0.3 dex, though 
there are cases where the uncertainty is particularly large;
for example, for HR 5459 ($\alpha$ Cen A) the \cite{edvardsson93} and 
the \cite{ng_} ages differ by a factor of two! Another example, HD 148816,
was assigned an age of 13.5 Gyr by \cite{edvardsson93}, but was assigned 
an age of only 5.3 Gyr by \cite{ng_}. Its V velocity of $-$220 km s$^{-1}$ 
suggests that HD 148816 is actually a halo star which would
be consistent with the larger age.
Nevertheless, we exclude HD 148816 from our analysis
because of the uncertainty; if included at the \cite{ng_} age, it increases 
the velocity dispersion in the log age = 0.7--0.8 bin from 42 km s$^{-1}$
to about 70 km s$^{-1}$. For this reason we excluded two additional stars
for which the \cite{ng_} and \cite{edvardsson93} ages differed by more than
0.3 dex.  Although the actual uncertainties (particularly
systematic errors) in the stellar ages are difficult to quantify, these
ages are undoubtedly a major improvement over the ages estimated during
the 1970s (e.g., \cite{wielen}) which typically used
mean ages for groups of stars of similar spectral types. 

Figure 1 shows the complete set of U, V, and W velocities for all stars in 
our sample (where U is defined as positive toward the Galactic center, 
V is positive in the direction of Galactic rotation, and W is positive 
toward the North Galactic Pole).  Figure 2 shows the total and U, V, W 
velocity dispersions derived from these data as a function of stellar age
(values are given in Table 1 for each age bin). The error bars 
are estimated from the scatter of the stars in each bin and so
represent the statistical error in the velocity dispersions only.  
A comparison of
the bottom panel of Figure 1 with the corresponding Figure 16b of 
\cite{edvardsson93} and the similar figure of \cite{freeman} shows that
the we find a slight, but noticeable reduction in the scatter in the
W velocities. A similar effect was noted by \cite{garnett} for the 
scatter in the age-metallicity relation compared to that found by
\cite{edvardsson93}. 

\subsection{An abrupt increase in velocity dispersion at an age 
of $9$ Gyrs}

We see from Figure 2 that there is a statistically significant and abrupt 
increase in all velocity components of the velocity dispersion at a 
stellar age of $9\pm 1$ Gyrs. If we use the size of the errorbars on the points 
to represent our uncertainty then the dispersion at 12 Gyrs is $\sim 2 
\sigma$ in the individual components and $\sim 3 \sigma$ in the total 
dispersion above the dispersions from 5-9 Gyrs.

Note that the data in Figure 2 show velocity dispersions computed
directly from the velocity data, with no weightings. \cite{wielen_}
advocated weighting the velocity differences by the amplitude of the
W component, appropriate for stars moving in a harmonic potential. 
A harmonic potential is appropriate for the low velocity dispersion
stars.  For stars which spend most of their orbit above the
bulk of the stellar population the weighting should be somewhat larger.
We show the results of weighting by $|W|$ in Figure 3.  In general, the
results are the same, with the exception that the jump in the U 
dispersion is reduced.  Nevertheless, the
jump in $\sigma$ at 9 Gyr remains significant.
We also note that our derived dispersions for
the oldest (thick disk) stars are in excellent agreement with the
values determined by \cite{chibabeers}, 
($\sigma_U$, $\sigma_V$, $\sigma_W$) = (46$\pm$4, 50$\pm$4, 35$\pm$3)
km s$^{-1}$.   Figure 4 shows the UV plane distribution of
the stars in 4 different age bins and illustrates that there
is little difference in the UV plane distribution of
the stars with ages between 2.5 and 5 Gyrs and those between 5 and 9 Gyrs.

We have been unable to imagine a selection effect that could lead to
kinematic biases in the sample.
Although the stellar selection criteria used by \cite{edvardsson93} 
causes the sample to contain a larger fraction of metal-poor stars than 
is actually the case in the solar neighborhood, 
contamination by halo stars is unlikely because the halo contributes
less than 1\% to the local stellar space density (\cite{schmidt75}).
Only one strong candidate for a halo star (HD~148816) is present in the sample.
The \cite{edvardsson93} 
sample may also be biased against selection of old, metal-rich stars,
but it is not clear that this would result in a bias against an old,
low velocity dispersion component if it exists. 

We therefore suspect that there is a genuine increase in the dispersion
at an age of about 9 Gyrs, which we associate with the onset of the thick
disk.  We can see from the figures that this increase 
happened on a timescale less than a few Gyrs.
An abrupt large increase is inconsistent with scattering models 
involving either molecular clouds or spiral arms. 
The only process discussed in the literature that could cause
such an abrupt and large increase in the velocity dispersion 
would be a merger that did not completely destroy the disk
of the Milky Way (e.g., \cite{toth}).
Numerical simulations 
(e.g.,  \cite{walker}; \cite{huang}; \cite{velazquez})
suggest that the Milky Way could have survived merging with
a smaller galaxy of up
to 1/3 of the Milky Way's mass (at that time).
Such a merger would result in significant thickening of the disk.
Based on the statistics of lopsided galaxies, \cite{zaritsky} 
estimate that minor mergers occur  in an $L_*$ galaxy 
about every 10 Gyrs.  This is consistent
with the possibility that the Milky Way suffered such an event $\sim 10$ 
Gyrs ago.
Cosmological models also predict that for an $L_*$ galaxy 
the merger rate would have been higher in the past 
(\cite{laceyandcole}).

\subsection{A quiescent disk between 3 and 9 Gyrs}

Our figures show that the velocity dispersion hardly 
increases between an age of 3 and 9 Gyrs rather
than smoothly increasing during this time 
(as previously proposed; \cite{wielen}).
\cite{freeman} suggested that the heating mechanisms 
become inefficient at $\sigma  \sim 30 $ km s$^{-1}$ and that
another process 
is required to explain the high
dispersions seen in the `thick disk'.

Heating solely from molecular clouds 
predicts $\sigma \propto t^{0.20}$ (\cite{lacey84}) and 
would be consistent
with the quiescent disk or minimal increase
in dispersion between 3 and 9 Gyrs ago.
However the observed dispersion ratio, $\sigma_z / \sigma_R$, is too low
to be consistent with this model. 
\cite{jenkins_} showed that this ratio  
can be set by the ratio of the heating rate (or radial
dispersion increase) caused by the two processes: 1) scattering from
spiral structure and 2) scattering from 
molecular clouds.  To match the observed dispersion ratio
the radial heating rate from transient spiral structure must be 
significantly larger than that from molecular clouds.
However, their model that best fit the observations of
the time predicted $\sigma \propto t^{0.5}$ 
which is inconsistent
with what we see between 3 and 9 Gyrs.

\cite{jenkins_} assumed that heating
from spiral structure was efficient for dispersions up to 70 km s$^{-1}$.
As pointed out by \cite{carlberg_} once the epicyclic amplitude
reaches a sizescale similar to that of the spiral arm wavenumbers,
heating is much less efficient and $\sigma \sim t^{0.2}$.
A star with such a large epicyclic amplitude would cross more 
than one spiral arm as it oscillates radially making it more difficult
for the spiral structure to perturb the star.
This lower exponent, $\alpha \sim 0.2$, is  
consistent with the nearly constant dispersions between 3 and 9 Gyrs
shown in Figure 2 and 3.

The simplest explanation for the observations is that both heating
processes are important (setting the ratio of the dispersions as modeled
by \cite{jenkins}) but that heating from spiral structure
becomes inefficient once the radial dispersion reaches about 30~km~s$^{-1}$.

The saturation point or maximum dispersion reached can be compared to the 
limit predicted by the observed wavenumber of spiral structure
in the Milky Way.
Heating from tightly wound spiral structure becomes inefficient when
$k a \sim 1$ where $k$ is the radial wavenumber
of the spiral structure and $a$ is the epicyclic amplitude 
(\cite{carlberg}).
The epicyclic amplitude is $a = \sqrt{2 E_R}/\kappa $
where $E_R$ is the energy in the epicycle, 
and $\kappa$ is the epicyclic frequency.
For a star with maximum radial velocity $V_R$, $E_R =  1/2 V_R^2$.  
$\kappa = \sqrt{2} v_c/R_0$ for a flat rotation curve
with circular velocity $v_c$ at a radius $R_0$ from the Galactic
Center.
%
%
%
Setting $E_R \sim \sigma_R^2$ (corresponding to a typical particle
with maximum radial velocity $\sim \sqrt{2} \sigma_R$),
we find that $k a \sim  1$ when $k R_0 \sim {v_c/\sigma_R}$.
For $v_c \sim 200$ km s$^{-1}$ 
and heating diminishing at $\sigma_R \sim 30$ km s$^{-1}$, 
consistent with the dispersions between 3 and 9 Gyrs, 
heating from transient spiral structure in the solar
neighborhood  becomes inefficient for $k R_0 \gtrsim 7$.  
This is fairly tightly wound; 
more tightly wound than that seen in the numerical simulations
of \cite{carlberg87}, $k R \sim 4.1$, which was
adopted as limits by \cite{jenkins_}.
This approximate limit can be directly compared to the winding number of
spiral structure observed in the Milky Way.

Tracers of spiral structure in the solar neighborhood
suggest that the spiral structure
is quite tightly wound (e.g., \cite{vallee}).
A two armed spiral is seen at K band in COBE/DIRBE data
with pitch angle in the range $p\sim 15.5^\circ$~--~$19^\circ$ (\cite{drimmel}),
giving $k R_0 = m \cot{p}$ in the range 5.8~--~7.2. 
This range is consistent with the approximate limit predicted
from the maximum radial dispersion reached between 3 and 9 Gyrs.
However, the tightly wound spiral structure
observed in the Milky Way is unlikely to efficiently heat 
the disk to the large velocities observed in the thick disk.
Spiral arms with a significantly higher pitch angle of 
$\sim 26^\circ$ (corresponding to $k R_0 \sim 4$) 
would be required to do this.
We find that the slow increase in stellar dispersions 
between 3 and 9 Gyrs observed
in Figure 2 is consistent with inefficient heating
from the tightly wound spiral structure 
observed in the Milky Way.
More detailed calculations are needed to more accurately determine
if the slow observed increase in dispersions can be predicted from a
model consistent with the observed spiral structure.

\cite{jenkins92_} found that the model cloud heating rate that  
fit the dispersions at ages greater than 8 Gyrs (data from \cite{wielen})
required an order of magnitude larger cloud number density than presently 
observed.  However, to fit the dispersions for the
stars younger than 8 Gyrs (data from \cite{stromgren}), 
he found that the heating required was consistent
with the present molecular cloud number density.
If an additional process is responsible for the largest
dispersions at ages greater than 8 Gyrs,  we expect that
there would be no discrepancy between the model heating
rates and observed dispersions.

The nearly constant dispersion between 3 and 9 Gyrs does not imply
that stars that are 8 Gyrs old were heated  
to 30 km s$^{-1}$ within 3 Gyrs. It may be that the oldest stars were
heated initially to a lower value (which might have been more
consistent with a less massive disk)
then subsequently were slowly heated as the disk
became more massive.  This would be consistent with a picture where
the Toomre stability parameter, $Q$, remained constant 
as the disk increased in mass.

\section{Is heating from spiral structure saturated in other galaxies?}

\def\spose#1{\hbox to 0pt{#1\hss}}
\def\proptoa{\mathrel{\spose{\lower 3pt\hbox{$\mathchar"218$}}
     \raise 2.0pt\hbox{$\propto$}}}

There are two approaches to testing any scenario for disk
heating.  Firstly the properties of stars in the Milky Way disk
can be more carefully studied and compared to evolution models.
Alternatively,
one can look for scaling relations betwen disk properties in other galaxies. 

We consider the possibility that heating from transient spiral structure
has saturated in other disk galaxies.  
Galaxies with spiral structure are expected to have a Toomre
stability parameter within the range $1<Q<2$ where 
$Q \approx { \sigma_R \kappa  \over \pi G \Sigma}$ (see \cite{B+T}).
This defines a relation between the radial velocity dispersion and quantities
that can be estimated from observations; the disk surface density
$\Sigma$ (which is the product of the surface
brightness and the mass to light ratio) and the rotation curve
(which determines $\kappa$).
If the disk lies at the maximum dispersion above
which heating from spiral structure ceases to be efficient,
then the winding number of the spiral structure is related to the 
radial dispersion by $k R \sim v_c / \sigma_R$ (see discussion above).  
If we combine this with the definition
for the Toomre $Q$ stability parameter,
we find that
\begin{equation}
k R \sim {v_c \kappa \over Q \pi G \Sigma}.
\end{equation}
If galaxies have radial dispersions at the saturation point 
where heating from spiral structure ceases to be efficient then 
we can relate the winding number of spiral structure to the underlying
disk surface density.

We now consider how the right-hand side of the above relation varies between
galaxies.  
Disk central surface brightnesses tend to be similar in many 
high surface brightness galaxies;
$\mu_{0,B} \sim 21.65$ mag arcsec$^{-2}$ at B band  
(\cite{freeman70}). 
At $R = 2.2 h$, where $h$ is the disk exponential scale length, 
(approximately the location of the solar neighborhood in the Milky Way
and where the rotational velocity is expected to reach a maximum)
the heating becomes inefficient when
\begin{equation}
k R \sim  4.0
\left({v_c \over 200 {\rm km/s}} \right)^2
\left({\gamma \over \sqrt{2}}\right)
\left({S(21.65 {\rm mag ''^{-2})}\over S_{0,B}}\right)
\left({5.0  M_\odot / L_{\odot,B} \over M/L_B }\right)
\left({3.0 {\rm kpc}\over h} \right)
\left({2.2 h \over R } \right)
\left({1.5  \over Q } \right)
\left({e^{r/h} \over e^{2.2}} \right)
\end{equation}
where $S(21.65 {\rm mag}''^{-2})$ corresponds to the surface brightness
(not in magnitudes)
which is equivalent to $21.65$ mag arcsec$^{-2}$ at B band.
Given typical disk parameters, the scaling relations given by 
equation 2 could be satisfied in many disk galaxies.
In other words,
radial dispersions could have reached maximum values where 
heating from transient spiral structure ceases to be efficient.

Based on scaling from the cosmological spin parameter,
we expect that the disk exponential scale length 
depends on the luminosity
$h \propto L^{1/3}$ with a large scatter (\cite{dejong}).
Using the Tully-Fisher relation, $L \propto v_c^4$, we find that
$k R \proptoa  L^{1/6}$ at saturation.
This implies that more massive galaxies, if they are near saturation,
should have slightly more tightly wound spiral
structure.  This is consistent with the the general trend
of the Hubble sequence; more tightly wound galaxies are of earlier-type.

Using observed galaxy surface brightness profiles (preferably at a wavelength
not strongly affected by dust and young stars such as K band) and
assumed or measured values for the mass-to-light ratio and $Q$, 
it is possible to estimate the value of $k R$ for a disk 
with a maximum radial dispersion.
If $k R$ estimated from equation (2)  
is greater than that actually observed from spiral
structure in the galaxy, then the stellar
radial dispersion should be increasing (spiral structure is
actively heating the disk).
If the estimated $k R$ is lower than that observed, 
then the observed spiral structure is incapable of causing the inferred 
high radial dispersion in the disk.  This would suggest that
another process, such as a merger or galaxy flyby
may have caused a stellar dispersion increase.
This kind of study could also provide
evidence for past mergers in nearby galaxies.

\section{Summary and Discussion}

\cite{freeman} and \cite{stromgren_} suspected that
the age-dispersion relation of solar neighborhood stars 
was not consistent with the smoothly increasing relation
proposed by \cite{wielen_}. 
In this paper we 
re-examine the age-velocity dispersion relation in the solar neighborhood
using improved stellar age estimates which are based on Hipparcos parallaxes
and recent stellar evolution calculations.
We find that there is an abrupt statistically significant jump in all velocity
components at age of about 9 Gyrs, confirming the statements made
by \cite{freeman} and \cite{stromgren_} and modeling
done by \cite{larsen}.
While the slow diffusion model proposed
by \cite{wielen_} is consistent with the theoretical models
for ages up to about 9 Gyrs, it is unlikely to explain the abrupt increase
in dispersions seen at this time.  The most likely explanation is
that the Milky Way suffered a minor merger that formed the thick disk.

The best theoretical modeling of the age-dispersion relation 
(e.g., \cite{jenkins92}) was able to fit the age-dispersion
relation proposed by \cite{wielen_} but was still inconsistent
with the observed number density of molecular clouds.
\cite{jenkins92_} noted that this problem was resolved
if he fitted only stars younger than 8 Gyrs.
As previously discussed in the literature (\cite{lacey84}, \cite{jenkins})
a model incorporating both scattering from molecular clouds
and spiral structure is required to match the ratio of the vertical
to planar velocity dispersions.
However, to be consistent with the nearly constant observed dispersions
between 3 and 9 Gyrs
a model with more tightly wound spiral structure than 
assumed by \cite{jenkins_} is required.  
In such a model, once the epicyclic amplitude
reaches a size scale similar to that of the spiral arm wavenumbers,
heating is much less efficient and the velocity dispersion increases
extremely slowly.
The limit given by the observed plateau in dispersion
is consistent with the tracers of spiral
structure in the Milky Way which find that the local
spiral structure is quite tightly wound (\cite{vallee}; \cite{drimmel}).

Since we find little evidence for an increase in velocity
dispersion for stars between 3 and 9 Gyrs old, 
disk heating must occur primarily within 3 Gyrs of birth. 
However, 3 Gyrs is only about 12 rotation periods at the solar circle.
This might suggest that a resonant heating model or bar induced
heating model might be more appropriate (e.g., as explored in \cite{dehnen})
to explain the rapid increase in dispersions in the solar neighborhood
rather than a random diffusive scattering model.
In such a model, the ratio of vertical to planar dispersions
would also be set by the relative strength of the planar heating
and vertical heating rates.  
%
If the ratio of vertical to planar dispersions and the Toomre 
stability parameter are nearly constant with radius
(as suggested by \cite{bottema}),
then we might expect that spiral structure is more 
important than the Galactic bar 
in heating the stellar disk in the solar neighborhood. 
Alternatively, the short timescale implies that local
heating rates could be highly sensitive to the recent history  
of spiral structure or the Galactic bar.  In this case
we might expect strong variations in the heating rates
and stellar dispersions as a function of radius.   

It is possible that many galaxies have 
radial velocity dispersions at 
the maximum value where spiral structure ceases efficient 
heating of the disk.  Using surface
brightness profiles and assumed or measured values for the
mass-to-light ratio and Toomre $Q$ parameter, we suggest 
that the winding number of spiral structure 
consistent with a disk at this particular dispersion 
can be predicted.
If the winding number of the observed spiral structure is 
above the predicted value, then the spiral structure
is incapable of heating the stars to their estimated
radial velocity dispersion.
By comparing predicted winding numbers with those observed
in a sample of galaxies spanning a range of properties 
the hypothesis that
disks tend to lie near their maximum dispersions can be tested.
This type of study could also provide evidence for
past mergers in other galaxies.
Indeed the large scatter in vertical scale heights seen
in edge-on galaxies and tendency of interacting
galaxies to have larger vertical scale heights (\cite{schwarzkopf})
suggests that galaxy interactions and mergers may be a dominant cause
of vertical heating in disk galaxies.

Much of the controversy concerning 
the age-dispersion relation is due to the difficulties
in measuring accurate ages for a large number of solar neighborhood
stars.  We expect that in the next few years, through more
comprehensive metallicity studies, and with improved stellar evolution tracks,
a larger number of stars from the Hipparcos catalog will have 
improved ages estimates.  The trends discussed in this paper
can then be more thoroughly tested with a larger sample.

\acknowledgments
We thank Caty Pilachowski for giving us a routine
originally written by Dave Soderblom for computing
U, V and W components.
Support for this work was provided by NASA through grants NAG5-7734,
NAG5-3359, and grants GO-07869.01-96A and GO-07886.01-96A
from the Space Telescope Institute, which is operated by the Association
of Universities for Research in Astronomy, Incorporated, under NASA
contract NAS5-26555.
We also acknowledge helpful correspondence with Andy Gould and Walter
Dehnen.

\clearpage


\clearpage

\begin{figure}
\vspace{16.0cm}
\includegraphics{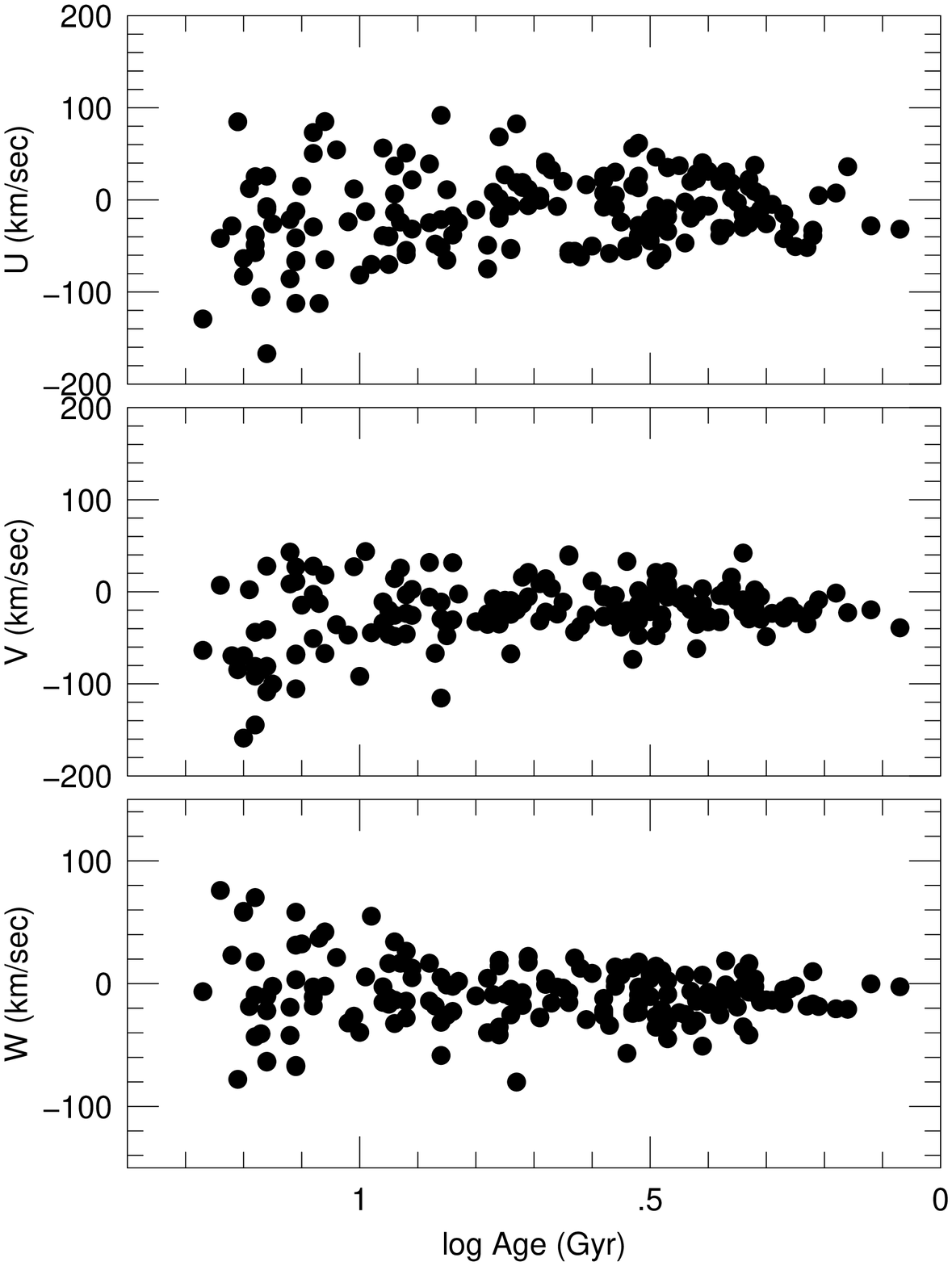}
\caption{Velocity components versus log age for the stellar sample of 
Edvardsson et al. (1993). The ages were taken from Ng \& Bertelli (1998)
as described in the text. The panels show the U, V, and W components 
for each star. Note the abrupt increase in the dispersion of each component
for stars older than 10 Gyr, and the lack of increase in velocity dispersion
between 3 and 9 Gyr. The bottom panel can be compared with Figure 31 of
Edvardsson et al. (1993) to see the effect of the revised ages on the
dispersion plot.}
\end{figure}

\begin{figure}
\vspace{16.0cm}
\includegraphics{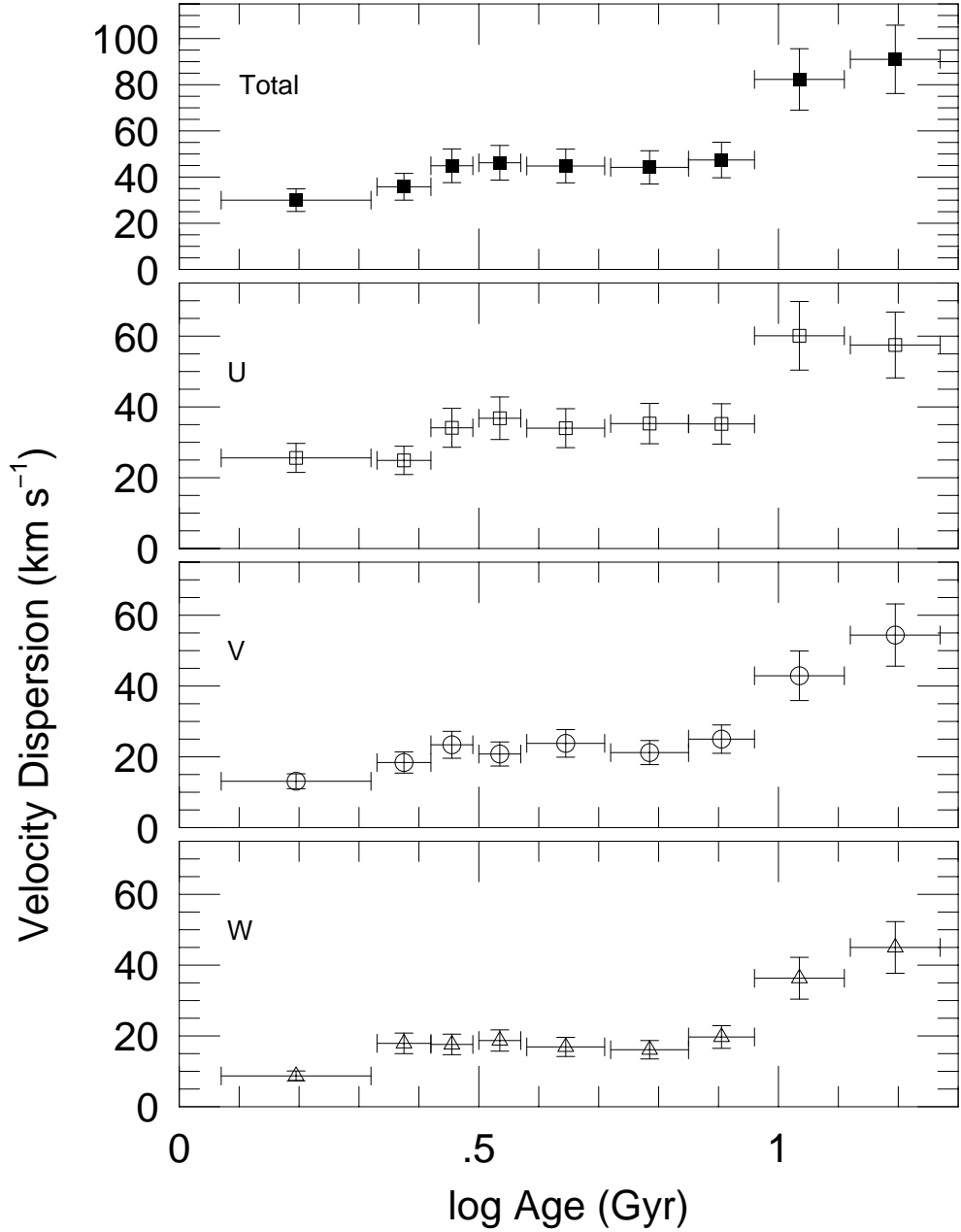}
\caption{Velocity dispersion versus log age for the data shown in Figure 1.
The points represent the unweighted dispersions. The top panel shows the
total dispersion 
$\sigma$ = ($\sigma_U^2$ + $\sigma_V^2$ + $\sigma_W^2$)$^{0.5}$.
Error bars reflect only the statistical uncertainty in velocity dispersions.
The dispersions are nearly constant between 3 and 9 Gyrs, but a significant
jump is seen at around 9 Gyrs.
}
\end{figure}

\begin{figure}
\vspace{16.0cm}
\includegraphics{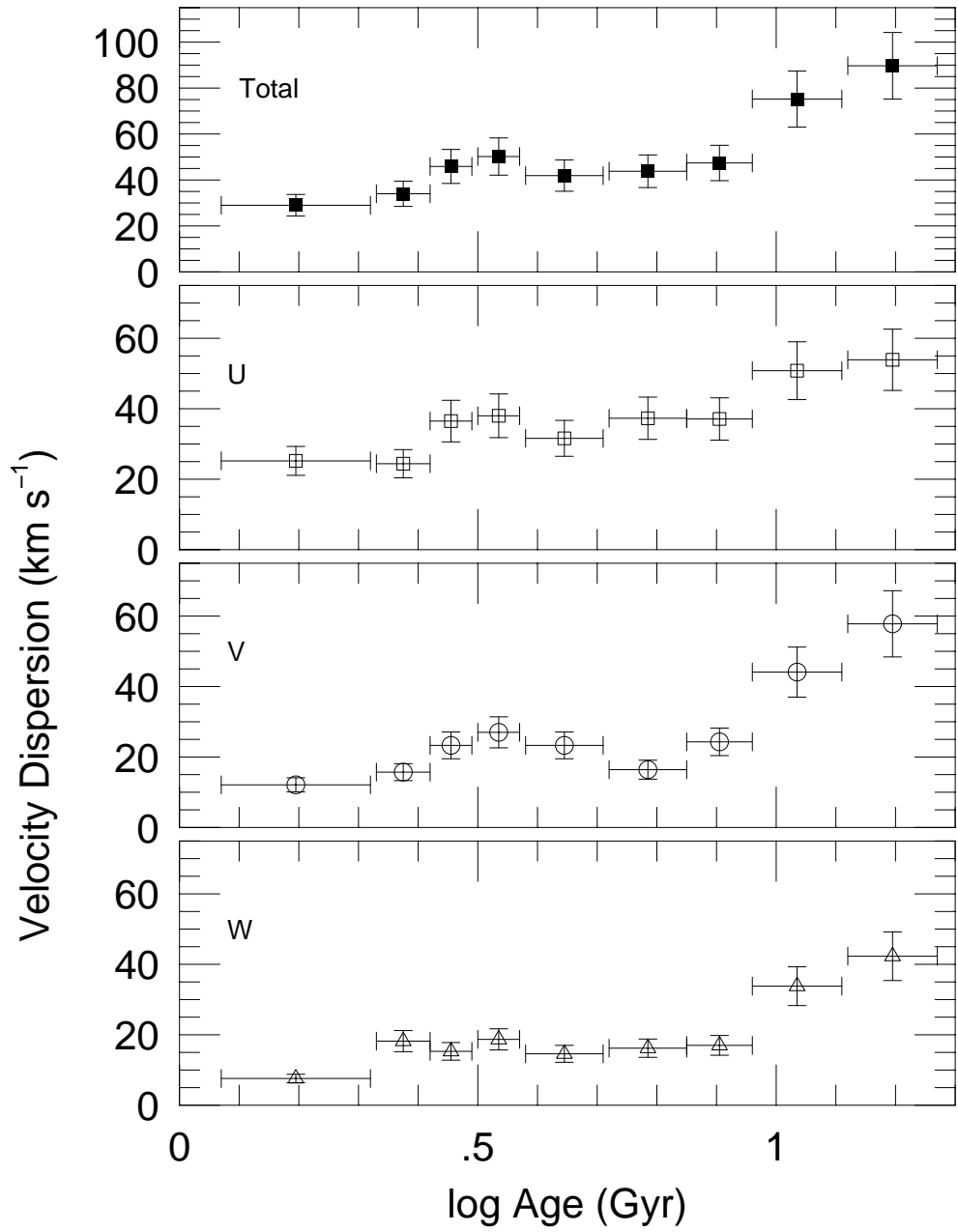}
\caption{Same as Figure 2, except that the velocity data have been weighted
by $|$W$|$ as per Wielen (1977). }
\end{figure}

\begin{figure}
\vspace{14.0cm}
\includegraphics{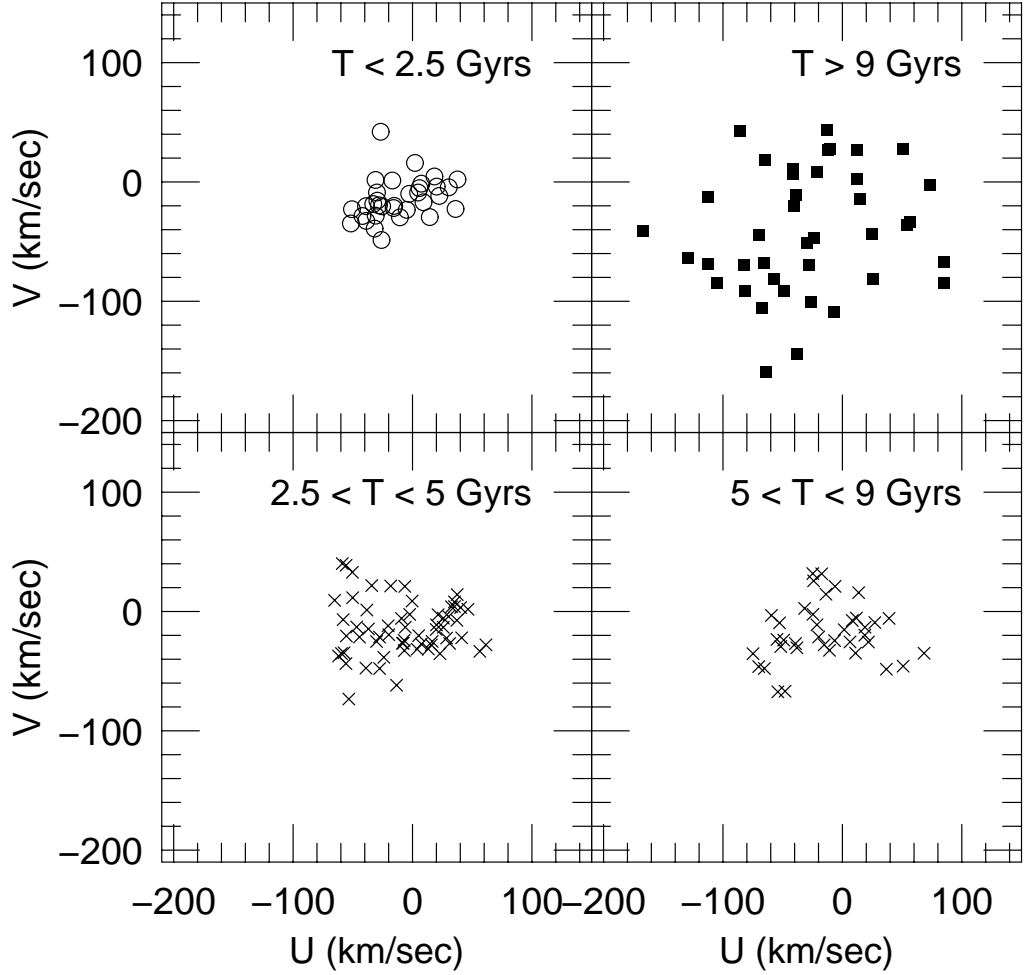}
\caption{A plot of the radial velocity component $U$ vs.~the tangential
velocity component $V$, for stars in different age bins. Note that the
distribution of velocities is very similar for stars between 
2.5 and 5 Gyrs and stars between 5 and 9 Gyrs. Stars with 
$t> 9$ Gyrs have a much larger dispersion.
In the $UV$ plane the distribution of the stars chosen by
Edvardsson et al.~(1993) 
is similar to that of the larger sample studied by \cite{dehnen_}.
}
\end{figure}

\clearpage

\begin{deluxetable}{lccccc}
\tablecolumns{6}
\tablewidth{0pc}
\tablecaption{Velocity Dispersions for Solar Neighborhood Stars}

\tablehead{ \colhead{Log t} & \colhead{$\sigma$(U)} & \colhead{$\sigma$(V)} 
& \colhead{$\sigma$(W)} & \colhead{$\sigma$(total)} & \colhead{N(stars)}\\   
\colhead{(Gyr)} & \colhead{(km s$^{-1}$)} & \colhead{(km s$^{-1}$)} & \colhead{(km s$^{-1}$)}
& \colhead{(km s$^{-1}$)}& \colhead{}  
}
\startdata
\cutinhead{Unweighted values}
0.07-0.32 & 26$\pm$4  & 13$\pm$2 &  9$\pm$1 & 30$\pm$5  & 19 \\ 
0.33-0.42 & 25$\pm$4  & 18$\pm$3 & 18$\pm$3 & 36$\pm$6  & 19 \\ 
0.42-0.49 & 34$\pm$6  & 23$\pm$4 & 18$\pm$3 & 45$\pm$7  & 19 \\ 
0.50-0.57 & 37$\pm$6  & 21$\pm$3 & 19$\pm$3 & 46$\pm$8  & 19 \\ 
0.58-0.71 & 34$\pm$6  & 24$\pm$4 & 17$\pm$3 & 45$\pm$7  & 19 \\ 
0.72-0.85 & 35$\pm$6  & 21$\pm$3 & 16$\pm$3 & 44$\pm$7  & 19 \\ 
0.85-0.96 & 35$\pm$6  & 25$\pm$4 & 20$\pm$3 & 47$\pm$8  & 19 \\ 
0.96-1.11 & 60$\pm$10 & 43$\pm$7 & 36$\pm$6 & 82$\pm$13 & 19 \\ 
1.12-1.27 & 58$\pm$9  & 54$\pm$9 & 45$\pm$7 & 91$\pm$15 & 19 \\ 
\cutinhead{Weighted by $\vert$W$\vert$}
0.07-0.32 & 25$\pm$4  & 12$\pm$2 &  8$\pm$1 & 29$\pm$5  & 19 \\ 
0.33-0.42 & 24$\pm$4  & 16$\pm$2 & 18$\pm$3 & 34$\pm$6  & 19 \\ 
0.42-0.49 & 36$\pm$6  & 23$\pm$4 & 15$\pm$2 & 46$\pm$7  & 19 \\ 
0.50-0.57 & 38$\pm$6  & 27$\pm$4 & 19$\pm$3 & 50$\pm$8  & 19 \\ 
0.58-0.71 & 32$\pm$5  & 23$\pm$4 & 15$\pm$2 & 42$\pm$7  & 19 \\ 
0.72-0.85 & 37$\pm$6  & 16$\pm$3 & 16$\pm$3 & 44$\pm$7  & 19 \\ 
0.85-0.96 & 37$\pm$6  & 24$\pm$4 & 17$\pm$3 & 47$\pm$8  & 19 \\ 
0.96-1.11 & 51$\pm$8  & 44$\pm$7 & 34$\pm$6 & 75$\pm$12 & 19 \\ 
1.12-1.27 & 54$\pm$9  & 58$\pm$9 & 42$\pm$7 & 90$\pm$14 & 19  
\enddata
\end{deluxetable} 

\end{document}